# Low Layer Functional Split Management in 5G and Beyond: Architecture and Self-adaptation

Jordi Pérez-Romero[1], Oriol Sallent[1], David Campoy[1], Antoni Gelonch[1], Xavier Gelabert[2], Bleron Klaiqi[2]

[1]*Universitat Politècnica de Catalunya (UPC), Barcelona, Spain*
[2]*Huawei Technologies Sweden AB, Kista, Sweden*
jordi.perez-romero@upc.edu, sallent@tsc.upc.edu, david.campoy.garcia@upc.edu, antoni.gelonch@upc.edu,
xavier.gelabert@huawei.com, bleron.klaiqi@huawei.com

*Abstract*—Radio Access Network (RAN) disaggregation is emerging as a key trend in beyond 5G, as it offers new opportunities for more flexible deployments and intelligent network management. A relevant problem in disaggregated RAN is the functional split selection, which dynamically decides which baseband (BB) functions of a base station are kept close to the radio units and which ones are centralized. In this context, this paper firstly presents an architectural framework for supporting this concept relying on the O-RAN architecture. Then, the paper analyzes how the functional split can be optimized to adapt to the different load conditions while minimizing energy costs.

## I. Introduction

The evolution of communication networks Beyond Fifth Generation (B5G) and towards Sixth Generation (6G) is expected to deal with the diverse and challenging requirements of innovative immersive and real-time vertical services [1][2]. At the same time, the current trend for disaggregated and softwarized RAN design for next generation mobile Base Stations (BSs), aligned with efforts such as the Open RAN (O-RAN) Alliance [3], offer new opportunities for a flexible deployment and intelligent network management.

A BS typically consists of a remote radio head (RRH), which performs all Radio Frequency (RF) processing functionality (e.g., filtering and power amplification), and a baseband unit (BBU), which provides the remaining necessary signal processing functions (e.g., orthogonal frequency division multiple access (OFDMA) processing, channel coding, digital modulation, etc.) and upper layers of the radio interface protocol stack (e.g., medium access control, radio link control, etc.). In early generations, an RRH and a BBU were jointly placed at a cell site. However, the current trend nowadays is to centralize some baseband (BB) functions towards data processing centres that are connected to the cell sites via the so-called fronthaul (FH) links. This enables the exploitation of resource pooling and statistical multiplexing gains among multiple cells, facilitates the introduction of collaborative techniques for different functions (e.g., interference coordination), and more efficiently handles the complex requirements of advanced features of B5G systems, such as the use of massive multiple input multiple output (MIMO).

Deciding the functional split (i.e., which BB functions are kept close to the radio units and which ones are centralized) embraces a trade-off between the centralization benefits and the fronthaul costs for carrying data between distributed antennas and the processing centers. In this respect, Fig. 1 shows a flexible RAN architecture that provides the freedom to resolve the above trade-offs at any desired operation point. The approach consists in allocating certain computing capabilities close to the cell site, where the so-called baseband low (BBL) functions are executed using the computational resources at a far edge site, and certain computing capabilities at a centralized location, where the so-called baseband high (BBH) functions are executed using the computational resources at a more central site (e.g. an edge site or at the cloud). As more functions are moved towards the BBH, higher computing efficiency and higher coordination gain in radio resources can be achieved at the expense of higher fronthaul bandwidth requirements and increased latency. Moving functions towards the BBL leads to the opposite behavior. In this context, enabling the split of functions between BBL and BBH to be changed over time is referred to as a dynamic functional split. A dynamic functional split may therefore exploit the most appropriate split to satisfy the traffic demands and quality requirements for the offered services, such as a high data rate or low latency. Due to varying mobile traffic demand over time, the ability to dynamically select the optimal functional split is crucial for an efficient usage of the fronthaul bandwidth and the baseband processing resources.

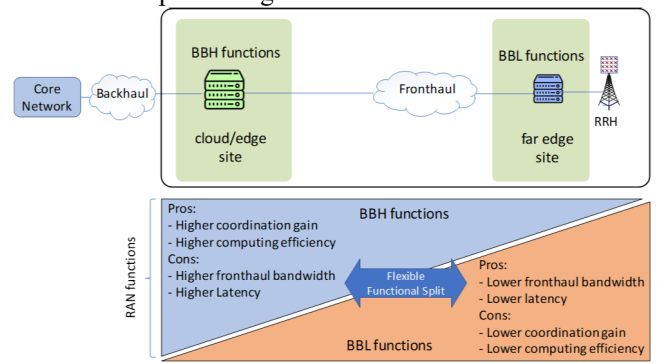

Fig. 1. Flexible functional split concept.

A lot of research has been conducted in recent years to shed light into the functional split problem, as reflected by some existing surveys that have summarized the key findings of different papers, research projects and industrial fora [4][5][6][7]. More recently, our paper [8] has focused on the low-layer functional splits that involve the BB processing functions at the physical (PHY) layer. The paper categorized the different algorithmic solutions of each BB function, gathering expressions of the computational complexity of each algorithm. This type of analysis is fundamental for properly assessing the implications of one or another functional split in terms of computational complexity requirements for the BBH and BBL. Based on the analysis of each BB function, the paper provided a system model to characterize the computational complexity and fronthaul bandwidth requirements of different functional splits.

Leveraging the work in [8], this paper deepens into two aspects of high relevance when targeting the practicality of dynamic functional split management. In this respect, a first contribution of the paper is the proposal of a novel architectural framework for implementing the dynamic functional split optimization, taking the O-RAN architecture as reference. Then, as a second contribution the paper analyzes how the functional split can be optimized to adapt to the load conditions in order to minimize the energy cost associated with the computational demand. This study considers in detail the computational requirements and fronthaul requirements arising from specific algorithmic solutions for the different BB functions based on the models from [8]. To the authors' best knowledge, this type of detailed analysis is missing in the open literature and constitutes one of the novelties of the paper, together with the proposed architecture.

The paper is organized as follows. Section II presents the considered architectural framework for enabling dynamic functional split optimization of the PHY layer functions. Then, Section III presents the adaptation of the functional split for a given optimization target and provides some results to assess the gains that can be achieved with respect to a fixed configuration. Finally, Section IV summarizes the main conclusions.

## II. ARCHITECTURAL FRAMEWORK

The Self-Adaptive Split Selection Application (SASSA) framework proposed in this paper targets the dynamic management of the BB functionality split between the BBH and BBL, responding to the time-varying nature of factors such as load demand, user density or service mixes. A high-level representation of the SASSA framework is illustrated in Fig. 2. It manages a Fifth Generation (5G) Radio Access Network (RAN) composed in general of multiple sites. Each site includes a number of sectors, each one with one or more 5G NR cells operating over a certain frequency and bandwidth. The computing resources are split between the BBH resources residing at a central location and the BBL resources, which host the PHY layer functionalities associated to the sectors of a site in accordance with the selected functional split. The FH network interconnects the different sites with the central location that hosts the BBH resources. The specific topology of the FH network will depend on the considered use case, but in general terms it will include a unit that multiplexes the FH information from the different sectors of a site, referred to in Fig. 2 as Site Switch Unit (SXU), and a unit that multiplexes the FH information coming from different sites, referred to in Fig. 2 as Central Switch Unit (CXU). An SXU and a CXU are interconnected by means of a FH link with a certain capacity.

The SASSA framework makes decisions on the selected functional split to use in a cell of a sector/site. These decisions will be based on monitoring information from the BBH, the BBL and the FH network and taking also into consideration the configuration parameters of a cell (e.g. number of physical resource blocks, operating frequency, number of antennas, configuration of the DeModulation Reference Signals (DM-RS) and Sounding Reference Signals (SRS), etc.). The monitoring of BBH and BBL includes not only the physical infrastructure resources (e.g. processors, memory, etc.) but also the performance of the PHY layer functions that run on them.

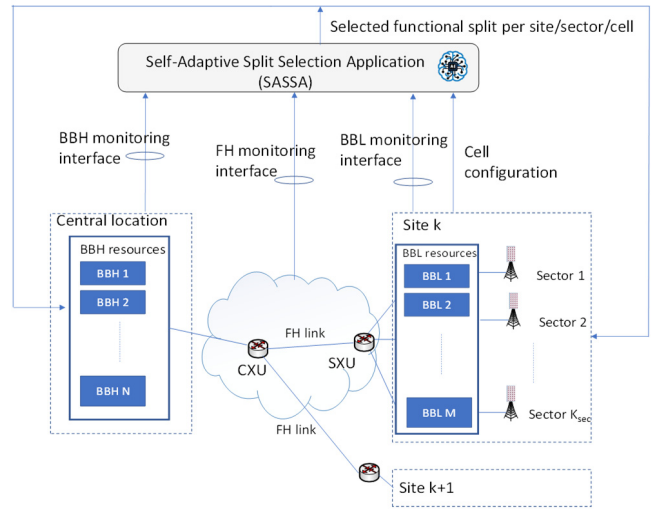

Fig. 2. High-level representation of the SASSA framework.

Based on the high-level representation of the SASSA framework, Fig. 3 presents a possible implementation aligned with the O-RAN architecture from [9], under the rationality that O-RAN is a significant industrial initiative aimed at specifying a disaggregated RAN, encompassing control and management functions designed to optimize the operation of such networks.

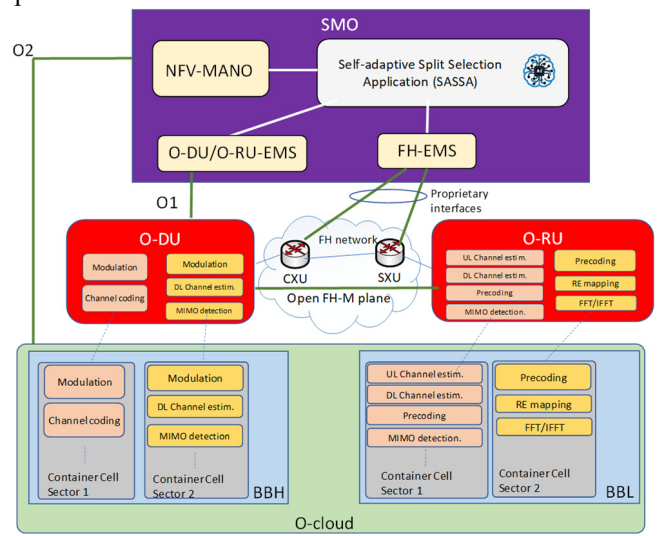

Fig. 3. Realisation of the SASSA framework based on the O-RAN architecture.

The O-RAN logical architecture is composed of disaggregated O-RAN functions and open interfaces as well as 3GPP interfaces. The O-RAN functions include the components for a disaggregated 5G NR gNB, namely the O-RAN Central Unit – Control Plane (O-CU-CP), the O-RAN Central Unit – User Plane (O-CU-UP), the O-RAN Distributed Unit (O-DU), and the O-RAN Radio Unit (O-RU). The SASSA framework considered in this paper, which addresses the low layer functional splits, involves the BB functionalities of the O-DU and the O-RU, as shown in Fig. 3. The O-RAN network functions can be implemented as Virtualized Network Functions (VNFs), e.g. in the form of Virtual Machines (VMs) or containers, or as Physical Network Functions (PNFs) utilizing customized hardware. In this respect, the support of virtualization in the O-RAN architecture is enabled by a cloud computing platform, denoted as O-Cloud.

The Service Management and Orchestration (SMO) system, on top of the O-RAN architecture, is responsible for the management of the O-RAN functions and the O-Cloud. From a wide network perspective, SMO functionalities encompass orchestration, management and automation procedures spanning across multiple network segments (e.g. core network, RAN, transport network, etc.). As seen in Fig. 3, the SASSA function that makes the decisions about the functional split in the different BB cells is associated with the SMO and can be implemented as an application (i.e. an rApp in O-RAN terminology) of the so-called non-real-time RAN Intelligent Controller (non-RT-RIC), which is one of the two intelligent controllers defined in the O-RAN architecture for closed-loop optimization control and operates in time scales above 1s.

The SASSA function is supported by other SMO functionalities that make use of different O-RAN interfaces. The O-DU and O-RU Element Management Systems (EMS) provide different management functionalities for the O-DU and O-RU nodes. In the proposed framework, they are used for monitoring the Performance Management (PM) counters to be exploited by the SASSA function when making decisions. The O-DU is managed through the O1 interface. Different PM counters from 3GPP TS 28.552 [10] can be collected. Among them, some relevant ones include the uplink (UL) and downlink (DL) traffic measurements per cell, the distributions of the used modulation and coding schemes and the Physical Resource Block (PRB) occupation per cell. In turn, the O-RU can be managed through the O-DU using the Open FH-M plane interface. In this case, relevant measurements are those related with the number of transferred packets between the O-DU and the O-RU.

The FH-EMS is in charge of managing the different components of the FH network, i.e. the SXU and CXU switches. This is outside the scope of O-RAN, which does not cover the management of the transport network. However, it is considered as a component of the SMO systems available to the operator for conducting management operations in the overall network. In this respect, the SASSA framework assumes that the different SXU and CXU switches that compose the FH network have the capability of providing PM measurements for their different ports, and that these measurements can be delivered to an FH-EMS management software through proprietary interfaces. Relevant FH monitoring metrics are those that allow capturing the occupation of the different links between the CXU and SXU units in a certain topology, including input and output data rates at the different ports.

Considering that the O-Cloud of O-RAN supports physical resources on multiple datacenters, the proposed architecture assumes that the O-Cloud encompasses the set of BBH and BBL resources available across the network. This enables the joint management of the virtualized functions deployed in these resources from the SMO through the O2 interface, which is defined between the SMO and the O-cloud. For this purpose, the SMO includes functionalities from the Network Function Virtualization (NFV) - MANagement and Orchestration (MANO) framework of ETSI [11] to enforce the decisions made by the SASSA function regarding the allocation of the virtualized BB functions of a cell to the BBH or the BBL in accordance with the selected functional split. The example represented in the lower part of Fig. 3 considers that the PHY layer functions of the different cells in different sites are containerized. In particular, for a given O-DU there are two containers at the BBH hosting respectively the functions of the cells 1 and 2, and other two containers at the BBL. It is worth mentioning that the current O-RAN architecture does not consider yet the virtualization of the O-RU [9]. This fits with the fact that O-RAN considers a fixed functional split between O-DU and O-RU function. In contrast, the proposed framework considers the possibility of virtualizing also the O-RU functions hosted at the BBL, because this facilitates the dynamic modification of the functional split done by the SASSA function.

III. SELF-ADAPTATION OF THE FUNCTIONAL SPLIT

Following the architectural framework described above, this section elaborates on the relevance of conducting the dynamic modification of the functional split to adapt to the traffic conditions in the different sectors and sites. To that end, the considered optimization target is presented first, followed by some illustrative results of the optimization.

A. *Optimisation target*

Let us consider a cell site that includes a number of sectors, each one with one 5G NR cell supporting massive MIMO. The PHY layer BB functions in a sector are those shown in Fig. 4. We consider that the traffic load offered to the different sectors varies along time and that the SASSA intends to optimize the selected functional split at each time for each one. The candidate functional splits that can be selected by the SASSA function are also shown in Fig. 4 and consider different splitting points denoted as (8,7a,7b,7c,7d, 6) where split 8 means that all the PHY layer functions are at the BBH, split 6 that all the PHY layer functions are at the BBL and splits 7a to 7d are ordered from more to less functions at the BBH.

Let us also denote as **X** the vector with the selected splits for each of the sectors in the cell site. Let also denote as $C_{BBH}(\mathbf{X})$ and $C_{BBL}(\mathbf{X})$ the actual computational requirements at the BBH and BBL, respectively, to support the configured splits in each sector. Note that the total computation required at the BBL and BBH for a given split can be derived by aggregating the computations required for each single PHY function.

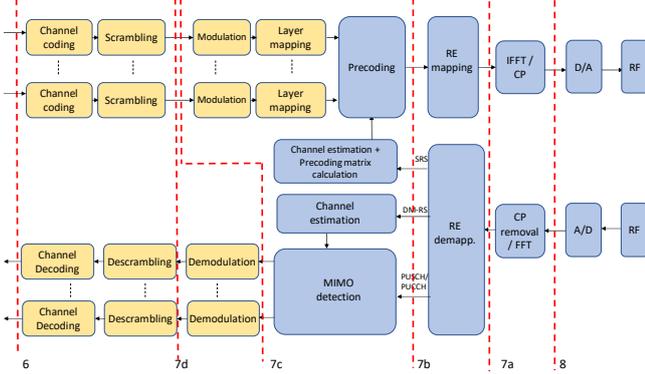

Fig. 4. PHY layer functions in DL transmission and UL reception and possible functional splits.

The functional split selection conducted by the SASSA function intends to minimize the energy consumption associated to the computational requirements at BBL and BBH. Moreover, considering that the computational cost can become a major constraint at remote sites [12], it is assumed that provisioning energy resources at the BBH is cheaper than provisioning the same resources at the BBL. Thus, let us define as $\varepsilon>1$ the ratio between the energy cost at the BBL and the corresponding cost at the BBH.

Regarding the fronthaul, it is assumed that the BBL of the site is connected to the BBH through an FH link with maximum capacity $R_{FH,max}$ (Gb/s). The traffic in the BBH to BBL direction (DL) and the traffic in the BBL to BBH direction (UL) use separate resources of the link, so that each direction has its own capacity $R_{FH,max}$. For a given combination $\mathbf{X}$ of functional splits and a certain traffic demand in each sector, let us denote as $R_{FH,DL}(\mathbf{X})$ and $R_{FH,UL}(\mathbf{X})$ the FH bandwidth requirement in the DL and UL, respectively.

With all the above considerations, the optimization problem consists in finding the combination of functional splits $\mathbf{X}$ that minimizes the total energy cost at the BBH and the BBL while maintaining the FH requirements below the FH capacity. This is mathematically expressed as:

$$\min_{\mathbf{X}} \left( C_{BBH}(\mathbf{X}) + \varepsilon \cdot C_{BBL}(\mathbf{X}) \right) \quad (1)$$

s.t.

$$R_{FH,DL}(\mathbf{X}) \leq R_{FH,max} \quad (2)$$

$$R_{FH,UL}(\mathbf{X}) \leq R_{FH,max} \quad (3)$$

### B. Evaluation scenario

For evaluation purposes we consider a site with three sectors, each one with a 5G NR cell operating with Time Division Duplex (TDD) and the parameters listed in Table I.

To characterize the BBH and BBL computational requirements and the UL and DL fronthaul bandwidth requirements we make use of the numerical model presented in our recent paper [8], where a detailed characterization of different algorithms for the BB functions and different functional splits was provided. Building on this study, the specific algorithms for each BB function are given in Table II. The table also plots the number of operations for one execution of each algorithm based on the cell parameters of Table I. The capacity of the FH link is $R_{FH,max}$=40 Gb/s and $\varepsilon$=2 is considered.

Regarding the variation of the traffic load in the cell site over time, Fig. 5 depicts the considered evolution over different periods of one day. For each period, the figure also indicates the percentage of load that corresponds to each one of the three sectors.

TABLE I. CELL PARAMETERS.

| Parameter | Value |
| --- | --- |
| Frequency | 3.5 GHz |
| Subcarrier spacing | $\Delta f$=30 kHz |
| Channel bandwidth and number of PRBs | 100 MHz, $N_{PRB}$=273 PRBs |
| Slot duration | $T_{slot}$=0.5 ms |
| Symbol duration | $T_s=T_{slot}/14$=35.7 μs |
| IFFT/FFT size | $N_{FFT}$=4096 |
| Number of time samples of cyclic prefix | $N_{CP}= N_{FFT}/14$=292 |
| Number of antennas at the BS | $B$=64 |
| Number of antennas at the UEs (equivalently number of layers) | $U$=16 (aggregate number for all UEs) |
| Length of training sequences for channel estimation | $L_{SRS}$=12, $L_{DM-RS}$=8 |
| Modulation and coding scheme | Assumed an average CQI in the cell corresponding to 64QAM ($m$=6 bits/symbol) and code rate $r$=666/1024. |
| TDD configuration | 3 DL slots, 1 special slot (with 10 DL symbols, 2 guard symbols and 2 UL symbols), 1 UL slot |
| Number of bits to encode an IQ sample | $n_{IQ}$=32 |
| Number of bits to encode a softbit at the output of the demodulator | $n_{soft}$=8 |
| Channel decoding parameters (see section IV.C.2 of [8]) | $BLER$=0.1, $I_{max}$=10, $d_c$=2, $n$=8424/$r$=12952, $d_v$=0.699, $E$=4528 |

TABLE II. CONSIDERED ALGORITHMS FOR EACH BB FUNCTION

| BB function | Algorithm | Operations |
| --- | --- | --- |
| FFT/IFFT | Radix-4 [13] | 49,152 |
| UL Channel Estimation | Beamspace Local Linear Minimum Mean Square Error (LMMSE) [14] | 297,984 |
| MIMO detection | Beamspace Local LMMSE [14] | 640 |
| DL channel estimation | Beam Space Channel Estimation [15] | 727,072 |
| Precoding matrix computation | Zero Forcing | 151,808 |
| Precoding | Matrix multiplication | 4,096 |
| Demodulation | Maximum Likelihood | 838 |
| Channel coding | Richardson Urbanke [16] | 12,952 |
| Channel decoding | Flooding [17] | 181,128 |

### C. Results

The optimum functional split per sector selected by the SASSA function in each period of time when minimizing the problem (1) subject to the FH bandwidth constraints (2)(3) in accordance with the considered traffic load variations is indicated in Fig. 5. This optimum functional split has been obtained for benchmarking purposes through an exhaustive search among all the possible combinations of splits after numerically computing the energy cost for every split. Results show how the optimum split varies for each sector depending on its load. For low loads more BB functions are centralized at the BBH using split 7b, while for higher loads some sectors tend to use split 7c by shifting some functions to the BBL. The rationality of this behavior is that, since the energy cost of the BBL is larger than that of the BBH, it is beneficial to centralize functions to the BBH by using lower splits such as 7b. However, these splits increase the required FH bandwidth, so they may not always fulfil the constraints (2)(3), and thus 7c may be needed in some cases.

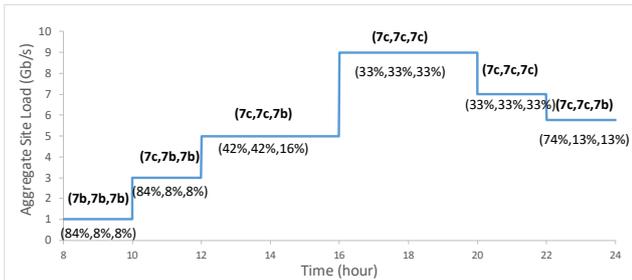

Fig. 5. Self-adaptation of the functional split per sector as a function of the aggregate site load.

Fig. 6 shows the evolution of the resulting energy cost, i.e. the minimum value of (1) measured in Giga Operations Per Second (GOPS), when selecting the optimum functional split in the different periods of the day. It is clearly observed that the energy cost follows a similar pattern like the aggregate site load of Fig. 5, reflecting that the computational requirements increase with the traffic. The reason is that most of the BB functions, like the MIMO precoding, the MIMO decoding or the channel estimation are executed only for the OFDMA symbols or PRBs that are occupied. Thus, increasing the traffic means that these functions need to execute more operations.

To quantify the improvements achieved through the adaptive split selection for each sector, Fig. 7 shows the evolution of the percentage difference in energy cost with respect to the optimum when using different fixed splits in all the sectors. The considered splits are 7c, 7d and 6 because these are the only ones that can be supported with the current FH capacity throughout the whole duration of the considered period. Instead, the splits 8 and 7a always demand a FH bandwidth that is higher than the capacity $R_{FH,max}$=40 Gb/s, while using split 7b in all sectors is only feasible in certain periods with low load. It is observed in Fig. 7 that having split 7c in all sectors is only optimum in the time period between 16h and 22h, while it becomes a suboptimum solution during the rest of the day, with differences to the optimum that vary between approximately 4% and 14%. The highest difference occurs for the first period between 8h and 10h. For this low load, as seen in Fig. 5, the optimum solution is having 7b in all sectors, thus moving all the precoding, MIMO decoding and channel estimation functions to the BBH, which reduces the required cost with respect to split 7c in which these functions are kept at the BBL. In turn, with the fixed splits 7d or 6 the differences with respect to the optimum are larger, varying between 20-25%, because in this case many of the most computationally demanding functions remain at the BBL.

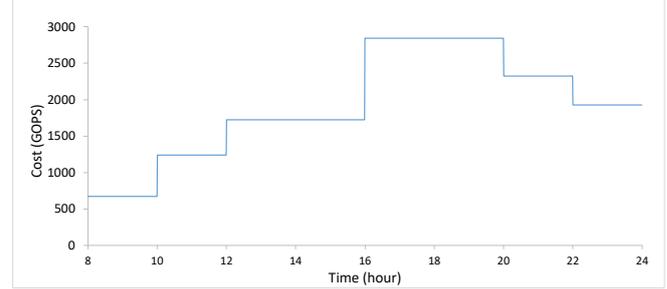

Fig. 6. Evolution of the energy cost with the optimum functional split selection.

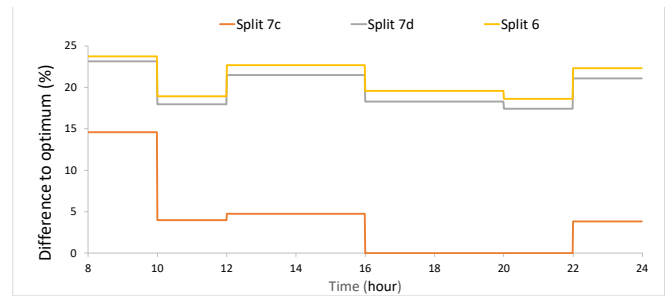

Fig. 7 Difference with respect to the optimum when using fixed splits.

Fig. 8 and Fig. 9 plot the FH bandwidth requirements in UL and DL, respectively, with the optimum adaptive split selection conducted by the SASSA framework and the different fixed splits. The limit $R_{FH,max}$=40 Gb/s is also indicated as a reference. Results illustrate the significant sensitivity of the fronthaul requirements with split 7b to traffic conditions, particularly in the DL, where they escalate rapidly from 30 Gb/s to nearly 300 Gb/s when the offered load is high. As a result, the use of this split in all the sectors is only feasible during the first period between 8h and 10h, while in the rest of the day the demanded bandwidth exceeds the limit of 40 Gb/s. In contrast, the FH requirements with split 7c are always kept below the maximum limit, thus making this option more adequate. In general, the higher splits 7d or 6 lead to a low utilization of the FH, as their demanded bandwidth uses to be much lower than the FH capacity. Then, the adaptive selection conducted by the SASSA framework is able to utilize as much FH bandwidth as possible while adhering to the constraints and accommodating varying bandwidth requirements based on the selected split in both UL and DL directions. Indeed, the capability of adapting the split on a per sector basis allows that, under certain situations, i.e. between 10h and 16h and between 22h and 24h, the split 7b can be selected for one or two sectors (see Fig. 5), which allows reducing the energy cost while at the same time increasing the FH utilization in relation to using a fixed split in all sectors.

Overall, the results presented in this section have demonstrated the necessity to have a flexible split selection strategy able to adapt the split on a per sector basis to traffic changing conditions. In some situations, traffic growth within a sector may increase the required fronthaul bandwidth and can cause the split combination in that site to become unfeasible, so that an adaptation to another combination is needed. In other situations, unbalanced traffic among sectors can be leveraged to offload low-traffic sectors using centralized splits, i.e. 7b, leading to heterogeneous optimum combinations.

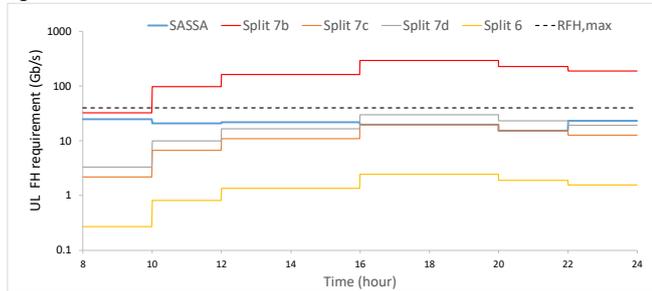

Fig. 8. UL fronthaul requirement comparison between fixed split strategy and adaptive split strategy.

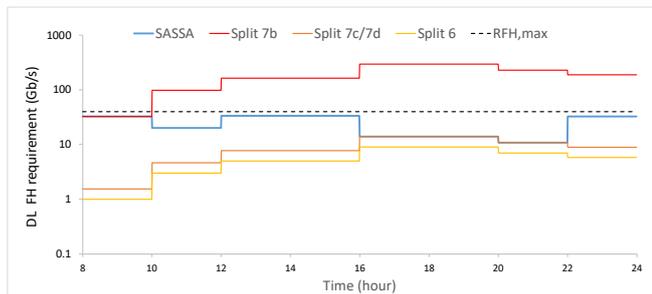

Fig. 9. DL fronthaul requirement comparison between fixed split strategy and adaptive split strategy.

## IV. CONCLUSIONS AND FUTURE WORK

This paper has addressed the functional split selection problem in a disaggregated RAN to decide the BB functions that have to be centralized and those that have to be kept close to the radio units. To that end, an architectural framework for self-adaptation of the functional split on a per cell basis has been presented. The framework is aligned with the functionalities of the O-RAN architecture. Specifically, the so-called Self-Adaptive Split Selection Application, which decides the selected functional split, is assumed to be part of the non-RT RIC at the SMO. This is supported by different measurements collected from the O-DU and O-RU nodes through the O1 and Open FH-M plane interfaces. Moreover, the O2 interface and the NFV-MANO are used to enforce the decisions on the allocation of BB functions in the BBH and BBL resources that are included as part of the O-cloud.

The benefits of the self-adaptation capability have been illustrated with an example with the target of optimizing the energy cost associated to the computation demands at the BBH and BBL. Results have shown that an adaptive strategy able to select the functional split on a per cell basis can lead to energy cost reductions between 4% and 25% with respect to the use of fixed splits in all cells. Moreover, it provides a better utilization of the available fronthaul capacity.

Based on the results presented here, future work intends to further assess the benefits achieved by the optimization considering other scenarios and to develop algorithmic solutions for conducting this optimization relying on the use of Artificial Intelligence and Machine Learning tools.

ACKNOWLEDGMENT

The work by J. Pérez-Romero, O. Sallent and D. Campoy has been partially funded by the Smart Networks and Services Joint Undertaking (SNS JU) under the European Union's Horizon Europe research and innovation programme under Grant Agreement No. 101096034 (VERGE project), by the Spanish Ministry of Science and Innovation MCIN/AEI/10.13039/501100011033 under ARTIST project (ref. PID2020-115104RB-I00) and by Huawei Technologies. The work by X. Gelabert and B. Klaiqi has been partially funded by the European Union's Horizon Europe research and innovation programme (HORIZON-MSCA-2021-DN-0) under the Marie Skłodowska-Curie grant agreement No. 101073265. The expressed views and opinions are, however, those of the authors only and may not reflect those of the European Union. Neither the European Union nor the granting authority can be held responsible for them.

REFERENCES

[1] K. Samdanis and T. Taleb, "The Road beyond 5G: A Vision and Insight of the Key Technologies," IEEE Network, vol. 34, no. 2, pp. 135-141, March/April 2020.

[2] W. Saad, M. Bennis, M. Chen, "A Vision of 6G Wireless Systems: Applications, Trends, Technologies, and Open Research Problems", IEEE Network, vol.34, 2020.

[3] A. Akman, et al, "O-RAN Minimum Viable Plan and Acceleration towards Commercialization", O-RAN Alliance White Paper, 2021.

[4] L. M. P. Larsen, A. Checko, and H. L. Christiansen, "A Survey of the Functional Splits Proposed for 5G Mobile Crosshaul Networks," IEEE Communications Surveys & Tutorials, vol. 21, no. 1, 2019.

[5] M. Peng, et al. "Recent Advances in Cloud Radio Access Networks: System Architectures, Key Techniques and Open Issues", IEEE Comm. Surveys & Tutorials, vol. 18, no. 3, 3rd Quarter 2016.

[6] I. A. Alimi, A. L. Teixeira, and P. Pereira Monteiro, "Toward an Efficient C-RAN Optical Fronthaul for the Future Networks: A Tutorial on Technologies, Requirements, Challenges, and Solutions," IEEE Communications Surveys & Tutorials, vol. 20, no. 1, 2018.

[7] B. Khan, N. Nidhi, H. OdetAlla, A. Flizikowski, A. Mihovska, J-F. Wagen, and F.J. Velez, "Survey on 5G Second Phase RAN Architectures and Functional splits", TechRxiv, Oct. 2022.

[8] J. Pérez-Romero, et al. " A Tutorial on the Characterisation and Modelling of Low Layer Functional Splits for Flexible Radio Access Networks in 5G and Beyond", IEEE Comm. Surveys & Tutorials, Vol.25, No. 4, Fourth Quarter, 2023.

[9] O-RAN Alliance, "O-RAN Architecture Description", O-RAN.WG1.OAD-R003-v11.00, February, 2024.

[10] 3GPP TS 28.552 v18.5.0, "Management and orchestration; 5G Performance measurements (Release 18)", December, 2023.

[11] ETSI GS NFV-MAN 001 v1.1.1, "Network Functions Virtualisation (NFV); Management and Orchestration", December, 2014.

[12] J. Liu, S. Zhou, J. Gong, Z. Niu, and S. Xu, "Graph-based framework for flexible baseband function splitting and placement in C-RAN," IEEE International Conference on Communications (ICC), Jun. 2015.

[13] M. Z. A. Khan and S. Qadeer, "A new variant of Radix-4 FFT," Thirteenth International Conference on Wireless and Optical Communications Networks (WOCN), 2016.

[14] M. Abdelghany, U. Madhow, A. Tölli, "Beamspace Local LMMSE: An Efficient Digital Backend for mmWave Massive MIMO", IEEE SPAWC, 2019.

[15] J. Shikida, K. Muraoka, N. Ishii, "Sparse Channel Estimation Using Multiple DFT Matrices for Massive MIMO Systems", IEEE 88th Vehicular Technology Conference (VTC-Fall), 2018.

[16] T. J. Richardson and R. L. Urbanke, "Efficient encoding of low-density parity-check codes," IEEE Transactions on Information Theory, vol. 47, no. 2, pp. 638–656, 2001. doi: 10.1109/18.910579.

[17] M. Benhayoun, M. Razi, A. Mansouri, A. Ahaitouf, "Low-Complexity LDPC Decoding Algorithm Based on Layered Vicinal Variable Node Scheduling", Modelling and Simulation in Engineering, 2022